\documentclass[11pt]{article}

\usepackage[utf8]{inputenc}
\usepackage[T1]{fontenc}
\usepackage[margin=1in]{geometry}
\usepackage{amsmath,amssymb,amsthm}
\usepackage{microtype} 
\usepackage{booktabs}
\usepackage{graphicx}
\usepackage{caption}
\usepackage{xcolor}
\usepackage[colorlinks=true,allcolors=blue!55!black]{hyperref}
\usepackage{natbib}
\usepackage{amsfonts}
\usepackage{bm}

\graphicspath{{figures/}}

\newcommand{\sigmoid}{\sigma}
\newcommand{\skill}{\theta}
\newcommand{\diff}{\delta}
\newcommand{\rider}{r}
\newcommand{\metric}{x}
\newcommand{\site}{s}

\title{\textbf{Inverse Suitability: Identifying Condition Difficulty and\\
Rider Skill from Behavioural Outcomes via\\ Continuous-Item Response Theory}}

\author{%
  Fabio Carucci\thanks{The method is developed and implemented within
  \emph{Goable}'s environmental calibration engine. The reproducibility artifact
  is the open command \texttt{inverse-suitability latent-demo}, released with the
  paper source at \url{https://github.com/goable-io/inverse-suitability}
  (Section~\ref{sec:repro}). Correspondence: \texttt{research@goable.io}.}\\[2pt]
  \emph{Goable} \\
  \texttt{https://goable.io}%
}
\date{\today}

\begin{document}
\maketitle

\begin{abstract}
Suitability scoring for outdoor activities---kitesurfing, paragliding, ski
touring---maps environmental conditions to a go/no-go verdict via expert-defined
curves. These curves conflate two distinct quantities: the intrinsic difficulty
of a condition and the skill of the person facing it. We introduce
\emph{Inverse Suitability}, a continuous-item Item Response Theory (IRT) model
that identifies both from behavioural outcomes alone. Each outcome is a triple
(rider $r$, condition metric $x$ at site $s$, binary outcome $y$); we model
$P(y{=}1) = \sigma\!\left(a\,(\theta_r - \delta(x,s))\right)$, where $\theta_r$ is
latent rider skill, $\delta(x,s)$ is a latent difficulty function anchored to a
physics-derived expert curve as its prior, and $a$ is a discrimination parameter.
The formulation is strictly more general than a single suitability curve, which
it recovers exactly when skill is integrated out under the population
distribution. Parameters are estimated by marginal maximum likelihood with
Gauss--Hermite quadrature; identification holds when the rider${\times}$condition
incidence graph is connected, with a documented single-curve fallback otherwise.
We validate via synthetic recovery: on a reference cohort ($80$ riders
$\times\,30$ outcomes) the model recovers latent skill at $r = 0.96$, locates the
difficulty minimum within $\pm 3$ units of ground truth, and improves held-out
Brier Skill Score by $+0.33$ over the expert-curve baseline. The recovered
difficulty function defines a measurable, site-level construct---an intrinsic
difficulty atlas---that existing meteorological observation networks do not
capture. All results reproduce from a single command on synthetic data, requiring
no proprietary observations.

\end{abstract}

\noindent\textbf{Keywords:} item response theory; latent-variable models;
suitability scoring; marginal maximum likelihood; environmental decision science;
skill identification.

\section{Introduction}
\label{sec:intro}

Operational decision support for outdoor activities rests on \emph{suitability
scoring}: a mapping from the environmental state---wind speed, wave height,
thermal lift, snow stability---to a bounded score or a go/no-go verdict. In
practice these mappings are expert-defined curves, one per activity and per
physical dimension, calibrated by domain specialists and, increasingly, refined
against observed outcomes. A kitesurfing wind curve peaks near a sweet spot and
falls off toward calm and storm; a ski-touring stability curve gates on
avalanche risk. The score is then a composition of such curves.

A single suitability curve, however, silently conflates two quantities that a
decision-maker needs to keep apart. The first is the \emph{intrinsic difficulty}
of a condition: how objectively demanding a given wind or sea state is for the
activity, independent of who is attempting it. The second is the \emph{skill} of
the person facing that condition. When a curve is fit to behavioural outcomes---
did the session go well, did the rider go out at all---it absorbs both: the
apparent ``suitability'' of $22$ knots at a given spot reflects not only how hard
$22$ knots is, but also the skill distribution of whoever happens to ride there.
Two spots with identical fitted suitability curves can therefore have radically
different intrinsic difficulty, masked by different clienteles. No existing
observation network---marine, atmospheric, cryospheric---measures the difficulty
term in isolation, because none carries the human-decision layer needed to
identify it.

\paragraph{The gap.} What is missing is a way to \emph{separate} intrinsic
condition difficulty from population skill using the outcome data that decision
platforms already collect, and to do so with an identified, interpretable model
rather than a black box. The methodological obstacle is that conditions are not
discrete test items: wind speed is a continuous axis, not a yes/no question, so
the classical psychometric machinery for separating item difficulty from person
ability does not apply off the shelf.

\paragraph{Contributions.} This paper makes three contributions.
\begin{enumerate}
  \item \textbf{Continuous-item IRT for environmental conditions.} We extend
    explanatory item response theory---which already lets item difficulty depend
    on covariates \citep{fischer1973,deboeck2004}---from discrete item features to
    a \emph{continuous physical driver}, modelling difficulty $\delta(x,s)$ as a
    smooth, physically anchored function of a metric $x$ at site $s$
    (Section~\ref{sec:model}). The item bank is the weather.
  \item \textbf{Identifiability and synthetic recovery.} We give the conditions
    under which rider skill and condition difficulty separate---a connected
    rider${\times}$condition incidence graph plus location/scale anchoring---and a
    deterministic marginal-maximum-likelihood estimator with a documented
    single-curve fallback when identification fails (Section~\ref{sec:ident}). We
    validate by synthetic recovery: on a reference cohort the model recovers
    latent skill at correlation $0.96$ and locates the difficulty minimum within
    $\pm 3$ units of ground truth (Section~\ref{sec:recovery}).
  \item \textbf{The difficulty atlas.} The recovered difficulty function defines a
    measurable, site-level construct---an \emph{intrinsic difficulty atlas}---that
    is anonymous, joinable to existing meteorological archives, and captures a
    quantity those archives do not (Section~\ref{sec:atlas}).
\end{enumerate}

\paragraph{Scope.} This is a methodology paper. Its validation is a synthetic
recovery study: cohorts are generated with known ground-truth skill and
difficulty, and the estimator is shown to recover them. We make \emph{no}
empirical claim on real riders; no proprietary observations are used anywhere,
and every number in the paper reproduces from a single command on synthetic data
(Section~\ref{sec:recovery}). Empirical validation on field cohorts is future
work (Section~\ref{sec:discussion}).

\section{Related work}
\label{sec:related}

Our contribution sits at the intersection of three literatures that have not
previously been connected: psychometric latent-trait models, meteorological
suitability indices, and latent-variable modelling in the environmental sciences.

\paragraph{Item Response Theory.} IRT is the standard framework in psychometrics
and educational measurement for separating the difficulty of a test item from the
ability of a test-taker, using only binary (correct/incorrect) responses
\citep{lord1968,rasch1960,embretson2000}. The two-parameter logistic model writes
the probability of a correct response as
$P(y{=}1) = \sigma\!\left(a_i(\theta_p - b_i)\right)$ for person ability
$\theta_p$, item difficulty $b_i$, and item discrimination $a_i$
\citep{birnbaum1968}. Parameters are classically estimated by marginal maximum
likelihood, integrating over the person-ability distribution with an EM algorithm
and Gauss--Hermite quadrature \citep{bock1981}; \citet{bakerkim2004} survey the
estimation techniques. The classical model treats items as \emph{discrete}, each
carrying its own scalar difficulty $b_i$ estimated from the responses to that
item. Two established lines of work already relax this by letting difficulty
depend on covariates: the linear logistic test model (LLTM), which writes item
difficulty as a \emph{linear} combination of item features \citep{fischer1973},
and the broader family of \emph{explanatory} item response models, which embed
item and person effects in a generalized linear (or nonlinear) mixed model
\citep{deboeck2004}. Our model is a continuous-item member of this explanatory
family. It extends LLTM/explanatory IRT along four axes: (i) the difficulty
covariate is a \emph{continuous physical driver} (e.g.\ wind speed), not a set of
discrete item features; (ii) difficulty is a \emph{smooth} function (spline or
Gaussian process) rather than LLTM's linear form; (iii) it is \emph{anchored to a
physics-derived curve} so its units are physically interpretable; and (iv) it is
applied to a new domain---environmental suitability---in which no two sessions
share an identical item, so the discrete-item view is not merely inconvenient but
inapplicable. Positioned this way, the contribution is an explicit extension of
explanatory IRT to a continuous environmental driver, not a claim that difficulty
has never been modelled as a function.

\paragraph{Suitability and climate indices.} Environmental decision science has a
long tradition of expert-defined indices that map weather to a rating. The Tourism
Climate Index \citep{mieczkowski1985} combines thermal comfort, precipitation,
sunshine, and wind into a single score; the Holiday Climate Index
\citep{scott2016} refines the weighting scheme, and second-generation indices
introduce climatic thresholds \citep{defreitas2008}. Activity-specific suitability
curves used in operational scoring engines are of the same family: expert-weighted
transformations of physical variables. These indices are, by construction,
population- and expert-level: they express how suitable a condition is
\emph{on average}, and none decomposes that rating into an intrinsic-difficulty
term and a skill term. Inverse Suitability takes such an expert curve as the
\emph{prior} for its difficulty function and then identifies the residual
structure that the single curve cannot express.

\paragraph{Latent-variable models in environmental science.} Latent-variable and
random-effects models are widely used in ecology and environmental statistics---
joint species distribution models built on latent factors \citep{warton2015},
occupancy models that separate a latent ``true state'' from an imperfect detection
process \citep{mackenzie2002}, and item-factor-analytic treatments of ordinal
environmental data. The closest in spirit are occupancy models: our skill term
plays a role analogous to a site- or observer-specific detection propensity,
separating a latent property of interest from the process that reveals it. What is
new relative to these is (i) anchoring the latent difficulty to a physics-derived
curve so that it is measured in interpretable units rather than recovered only up
to an arbitrary scale, and (ii) treating the difficulty as a smooth function of a
continuous physical driver rather than a per-site constant.

\paragraph{Forecast verification.} Our promotion metric, the Brier Skill Score,
is standard in forecast verification \citep{brier1950}: it measures the
mean-squared-error improvement of a probabilistic prediction over a reference. We
use it to test whether skill-conditioning earns its added complexity against the
single-curve baseline (Section~\ref{sec:recovery}).

\section{The model}
\label{sec:model}

\subsection{Outcomes and the link function}

The unit of data is an outcome triple $(\rider, \metric, y)$, optionally indexed
by a site $\site$: rider $\rider$ experienced a condition summarised by the metric
value $\metric$ (for example, wind speed) at site $\site$ and produced a binary
outcome $y \in \{0,1\}$---a good session versus not, or a go versus no-go. We
model the outcome probability with a two-parameter latent factorisation,
\begin{equation}
  \label{eq:model}
  P(y{=}1 \mid \rider, \metric, \site)
    = \sigmoid\!\Big( a(\metric,\site)\,\big(\skill_\rider - \diff(\metric,\site)\big) \Big),
  \qquad \sigmoid(z) = \frac{1}{1+e^{-z}},
\end{equation}
with three latent objects:
\begin{itemize}
  \item $\skill_\rider \in \mathbb{R}$ --- the \emph{latent skill} of rider
    $\rider$, one scalar per rider (per pseudonym), identified up to a
    location/scale convention (Section~\ref{sec:ident});
  \item $\diff(\metric,\site)$ --- the \emph{latent difficulty} of the condition:
    how objectively hard the metric value $\metric$ is at site $\site$, a smooth
    function of $\metric$;
  \item $a(\metric,\site) > 0$ --- the \emph{discrimination}: how sharply the
    condition separates riders of different skill. In the Phase-1 form used
    throughout this paper, $a$ is a single scalar.
\end{itemize}
Equation~\eqref{eq:model} is the two-parameter logistic IRT model with the item
difficulty $b_i$ replaced by a function $\diff(\metric,\site)$ of a continuous
driver. Higher skill or lower difficulty raises the success probability; the two
enter only through their difference $\skill_\rider - \diff(\metric,\site)$, a fact
with consequences for identification (Section~\ref{sec:ident}).

\subsection{Difficulty as a continuous item}

Classical IRT difficulty is a per-item scalar because test items are discrete.
Environmental conditions are not: the metric $\metric$ is a continuum, so we model
$\diff(\metric,\site)$ as a smooth function---a spline (or, with full Bayesian
inference, a Gaussian process) over $\metric$---rather than a collection of
independent per-condition constants. Two properties make this well-posed. First,
smoothness lets neighbouring conditions share statistical strength, so a condition
observed only a few times still receives a sensible difficulty. Second, the
difficulty function is \emph{anchored to the physics-derived expert curve as its
prior}, which serves two distinct purposes: it fixes the difficulty scale in
interpretable, physical units, and it acts as a shrinkage prior toward which
sparsely observed conditions regress. This anchoring is about \emph{units and
regularisation}, not identification: $\diff$ is already identified once the skill
distribution is standardised (Section~\ref{sec:ident}); without the anchor it
would still be identified, but expressed only on an arbitrary scale.

\subsection{Recovering the single suitability curve as a marginal}
\label{sec:marginal}

The model in \eqref{eq:model} is strictly more general than today's single
suitability curve, and reduces to it exactly. Let $p(\skill)$ be the population
skill distribution. The single suitability curve is the marginal of
\eqref{eq:model} over that population,
\begin{equation}
  \label{eq:marginal}
  \mathrm{curve}(\metric)
    = \mathbb{E}_{\skill}\!\left[\, \sigmoid\!\big(a(\metric)(\skill - \diff(\metric))\big)\,\right]
    = \int \sigmoid\!\big(a(\metric)(\skill - \diff(\metric))\big)\, p(\skill)\, d\skill .
\end{equation}
Integrating skill out recovers precisely the population-level curve that a
conventional calibrator would fit. This is not merely an asymptotic
correspondence; in our implementation it is a regression test---the fitted
model's marginal is checked against a single-curve fit on the same data
(Section~\ref{sec:recovery}, the \emph{marginal matches single curve} gate).
Consequently, adopting Inverse Suitability adds a dimension to an existing
scoring engine; it does not fork it. Where a cell lacks the data to identify the
latent structure, the model falls back to \eqref{eq:marginal}, which is the
behaviour the engine already has.

\subsection{Skill-conditioned scoring}

For a chosen skill level $\skill$---for instance a population quantile mapped to
beginner, intermediate, or expert---the skill-conditioned suitability curve is
$\metric \mapsto \sigmoid\!\big(a(\metric)(\skill - \diff(\metric))\big)$. This is
the per-level scoring surface: the same condition is scored differently for a
beginner and an expert, while the underlying difficulty $\diff$ is invariant
(Figure~\ref{fig:atlas}). The skill level is treated as an explicit input, not
inferred from identity---a boundary we return to in Section~\ref{sec:governance}.

\section{Identification and estimation}
\label{sec:ident}

Separating skill from difficulty is the make-or-break of the method. Three
obstacles stand in the way, each with an explicit remedy; the estimator is
otherwise the classical IRT one.

\subsection{Scale indeterminacy}

Because \eqref{eq:model} depends on skill and difficulty only through the
difference $\skill_\rider - \diff(\metric,\site)$, adding a constant $c$ to every
skill and to every difficulty leaves all outcome probabilities unchanged. Skill
and difficulty are therefore identified only up to a common shift (and, with a
free discrimination, a common scale). We use two devices with \emph{distinct
roles}---one that secures identification, one that fixes interpretable units:
\begin{enumerate}
  \item \textbf{Location/scale on skill (identification).} We fix the population
    skill distribution to $\mathbb{E}[\skill] = 0$ and
    $\mathrm{Var}[\skill] = 1$---the standard IRT convention. This removes the
    shift-and-scale freedom and is what makes $\skill$ and $\diff$ separately
    identified, not merely their difference.
  \item \textbf{Physical anchor on difficulty (units and prior).} Given
    identification, $\diff$ is still expressed on an arbitrary scale. Anchoring it
    to the physics-derived expert curve through its prior fixes the difficulty
    origin and units to interpretable, physical quantities, and simultaneously
    acts as a shrinkage prior for sparsely observed conditions
    (Section~\ref{sec:sep}). It is not what identifies the model.
\end{enumerate}

\subsection{The data condition: a connected incidence graph}
\label{sec:graph}

Identification also requires the right \emph{coverage}. Form the bipartite
incidence graph whose two vertex classes are riders and condition bins, with an
edge whenever a rider is observed in a bin. Skill and difficulty separate only
when this graph is \emph{connected}: riders must be observed across varied
conditions, and conditions must be experienced by varied riders. The failure mode
is intuitive---a rider who only ever goes out in easy conditions has a skill that
is confounded with ``always succeeds,'' because nothing in the data forces the two
apart. Before fitting, the estimator checks
\begin{itemize}
  \item connectivity of the rider${\times}$condition incidence graph;
  \item at least $K$ distinct condition bins per rider;
  \item at least $M$ riders per condition bin.
\end{itemize}
When these fail, the latent model \emph{auto-disables} and the estimator falls
back to the single-curve marginal of Section~\ref{sec:marginal}. This is a
deliberate refusal to over-claim: below the identification threshold, the honest
output is the population curve, not a spuriously ``personalised'' one. It also
makes the activation gate strictly stronger than a raw sample-size threshold---
volume alone does not identify skill; repeated cross-condition observation does.

\subsection{Separation and sparsity}
\label{sec:sep}

Two finite-sample pathologies remain. A rider with an all-success record drives
$\skill \to +\infty$ (the classical separation problem in logistic models); a
condition bin observed too rarely has an under-determined difficulty. Both are
handled by shrinkage: rare skills shrink toward the population prior, and sparse
difficulties shrink toward the expert-curve anchor. This keeps every estimate
finite and well-defined without introducing new machinery beyond the priors
already stated.

\subsection{Estimator: marginal maximum likelihood}
\label{sec:mml}

We estimate by \emph{marginal maximum likelihood} via EM, the classical IRT
estimator \citep{bock1981}. The skill of each rider is a latent variable; the
marginal likelihood of the observed outcomes integrates it out against the
population distribution $p(\skill) = \mathcal{N}(0,1)$. The E-step evaluates this
integral by Gauss--Hermite quadrature \citep{golub1969}; the M-step maximises the
difficulty function $\diff$ and the discrimination $a$ given the current
quadrature weights. The procedure is deterministic---same data and same seed give
the same result---which matches the reproducibility posture of
Section~\ref{sec:governance} and adds no heavy sampling dependency.

Full Bayesian estimation (for example Hamiltonian Monte Carlo with a Gaussian
process difficulty) yields richer posterior uncertainty bands and is a documented
upgrade; it is not required for the point-identification and recovery results of
this paper, and we treat it as future work (Section~\ref{sec:discussion}). The
per-rider skill posteriors and the difficulty function that the estimator returns
are exactly the objects validated next.

\section{Synthetic recovery experiment}
\label{sec:recovery}

%
\newcommand{\thetaRecoveryCorr}{0.96}
\newcommand{\thetaRecoveryCorrFull}{0.9638}
\newcommand{\difficultyMinKnots}{16.7}
\newcommand{\discriminationA}{2.16}
\newcommand{\heldOutBSS}{0.329}
\newcommand{\heldOutBSSsigned}{+0.33}
\newcommand{\nTrain}{2,400}
\begin{table}[t]
  \centering
  \caption{Synthetic recovery on the reference cohort ($80$ riders $\times\,30$
  paired outcomes, $n_{\text{train}} = 2,400$). Every value is recovered
  from behavioural outcomes alone by marginal maximum likelihood; none is supplied
  to the fit. The data-generating difficulty is minimised at the ground-truth
  optimum $x^\star = 18$~kn with true discrimination $a_{\text{true}} = 2.0$
  (stated in the text, not fit here). Reproduced by a single command
  (\texttt{inverse-suitability latent-demo}); see Appendix.}
  \label{tab:recovery}
  \begin{tabular}{@{}llr@{}}
    \toprule
    Quantity & Symbol & Recovered \\
    \midrule
    Latent-skill recovery (Pearson) & $\mathrm{corr}(\theta,\hat\theta)$ & 0.964 \\
    Difficulty-minimum location     & $\hat{x}^\star$ (kn)               & 16.7 \\
    Discrimination                  & $\hat{a}$                          & 2.16 \\
    Held-out Brier skill score      & $\mathrm{BSS}$ vs.\ single curve   & $+0.329$ \\
    Marginal matches single curve   & --                                    & \checkmark \\
    Identification gate passed      & --                                    & \checkmark \\
    \bottomrule
  \end{tabular}
\end{table}

Because no real cohort is used, the validity of the method rests on a
\emph{synthetic recovery} study: we generate data from a known ground truth, fit
the model, and measure how well it recovers that truth. This is the standard
validation for a methods paper, and it is the same role a simulation study plays
throughout statistics---it isolates the estimator's behaviour from the confounds
of any particular dataset.

\subsection{Data-generating process}
\label{sec:dgp}

The generator draws a two-factor cohort with \emph{known} skill and difficulty.
Ground-truth skills are $\skill_\rider \sim \mathcal{N}(0,1)$, one per rider. The
ground-truth difficulty is a parabola in difficulty space,
\begin{equation}
  \label{eq:dgp-delta}
  \diff(\metric) = \left(\frac{\metric - \metric^\star}{w}\right)^{2} + f,
  \qquad \metric^\star = 18~\text{kn},\; w = 8,\; f = -1,
\end{equation}
so that difficulty is \emph{minimised at the ground-truth optimum
$\metric^\star = 18$~knots} and rises symmetrically toward calm and storm---a
single-peaked suitability profile once passed through the link. Each rider is
observed across conditions drawn uniformly over $[3, 35]$ knots (guaranteeing a
connected incidence graph, Section~\ref{sec:graph}), and outcomes are Bernoulli
draws from \eqref{eq:model} with \emph{true discrimination}
$a_{\text{true}} = 2.0$. The reference cohort has $80$ riders $\times\,30$
outcomes each. These parameters---$\metric^\star = 18$~kn, $a_{\text{true}} = 2.0$,
and the cohort shape---are the ground truth supplied to the \emph{generator};
they are never supplied to the \emph{fit}. The recovered quantities reported below
are what the estimator infers from the outcomes alone.

\subsection{Metrics and results}

We assess recovery on three axes: (i) latent-skill recovery, the Pearson
correlation between the true $\skill_\rider$ and the estimated $\hat\skill_\rider$;
(ii) difficulty-shape recovery, summarised by the estimated location of the
difficulty minimum $\hat\metric^\star = \arg\min_\metric \hat\diff(\metric)$,
which should land near the true $18$~kn; and (iii) predictive gain, the held-out
Brier Skill Score of the skill-conditioned model against the single-curve
baseline under a stratified $80/20$ split. We also verify the backward-compatible
identity of Section~\ref{sec:marginal}---that integrating skill out of the fitted
model matches a single-curve fit---and that the identification gate of
Section~\ref{sec:graph} passes on this (deliberately connected) cohort.

Table~\ref{tab:recovery} reports the outcome. The model recovers latent skill at
correlation $\thetaRecoveryCorr$ against ground truth; it locates the difficulty
minimum at $\hat\metric^\star = \difficultyMinKnots$~knots, within $\pm 3$ knots of
the true $18$-knot optimum; and it improves the held-out Brier Skill Score by
$\heldOutBSSsigned$ over the single-curve baseline, confirming that
skill-conditioning earns its added complexity out of sample rather than merely
fitting the training data more flexibly. The recovered discrimination is
$\hat{a} = \discriminationA$ (true $a_{\text{true}} = 2.0$), the marginal matches
the single-curve fit, and the identification gate passes, all on
$n_{\text{train}} = \nTrain$ training outcomes.

\subsection{Reproducibility}
\label{sec:repro}

The experiment is a single command. Every value in Table~\ref{tab:recovery} is
emitted by
\begin{quote}
  \verb|uv run inverse-suitability latent-demo --out DIR|
\end{quote}
which fits the reference cohort and writes a \texttt{stats.json}; the paper's
table and inline figures are rendered directly from that file, so the reported
numbers cannot drift from the code (Appendix~\ref{app:repro}). The run is
deterministic (fixed cohort and fit seeds), completes in well under a minute on a
laptop, touches no proprietary data, and requires no network access. The
data-generating process is fully specified by \eqref{eq:dgp-delta} and
Section~\ref{sec:dgp}; a reviewer can regenerate the cohort and re-fit
independently.

\section{The difficulty atlas}
\label{sec:atlas}

The estimator returns two kinds of object. One is per-rider: the skill posteriors
$\hat\skill_\rider$. The other is a property of the environment: the difficulty
function $\hat\diff(\metric,\site)$ itself. Collected across sites and dimensions,
the latter forms what we call the \emph{intrinsic difficulty atlas}---a
site-level, anonymous map of how hard each condition is, net of who rides it.

\begin{figure}[t]
  \centering
  \includegraphics[width=0.86\linewidth]{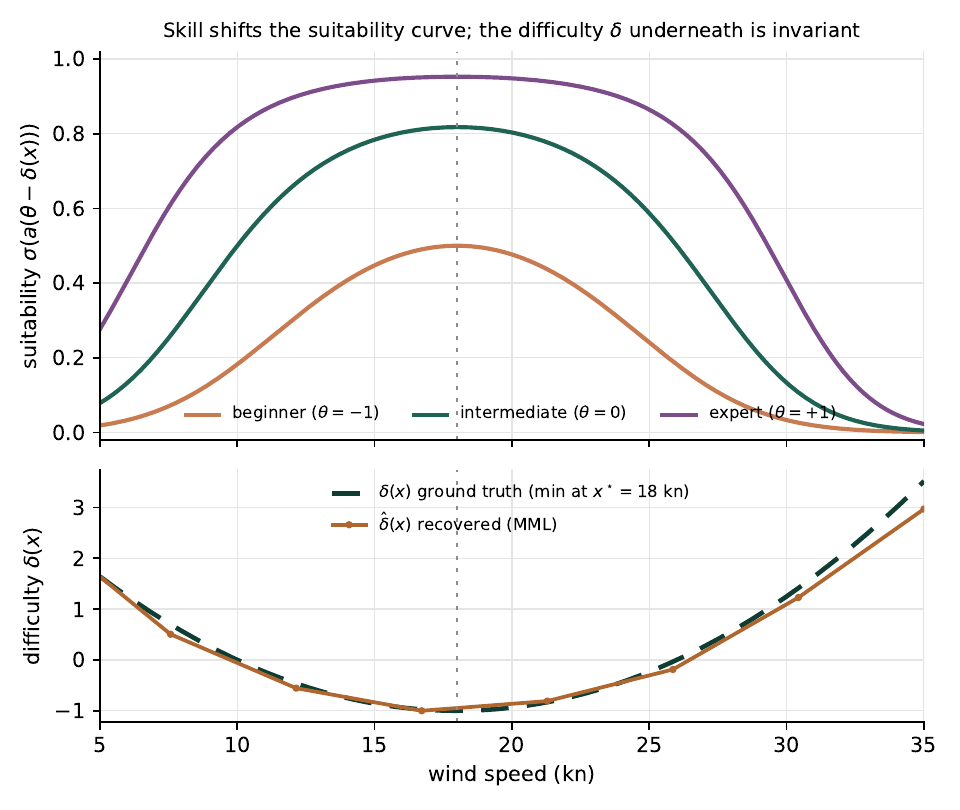}
  \caption{The construct the atlas measures. \emph{Bottom:} the intrinsic
  difficulty function $\diff(\metric)$ (dashed), minimised at the ground-truth
  optimum $\metric^\star = 18$~kn; the recovered $\hat\diff(\metric)$ is overlaid
  where available. \emph{Top:} three skill-conditioned suitability curves
  $\sigmoid(a(\skill - \diff(\metric)))$ for beginner ($\skill{=}{-}1$),
  intermediate ($\skill{=}0$), and expert ($\skill{=}{+}1$). Skill shifts the
  suitability curve while the difficulty beneath it is invariant: two operators
  whose spots share the same $\diff$ but differ in clientele read different
  suitability surfaces from the \emph{same} underlying difficulty. That
  separation is what no meteorological observation network records. Figure
  generated from code (Appendix~\ref{app:repro}).}
  \label{fig:atlas}
\end{figure}

\subsection{Why the atlas is new}

Existing observation networks---marine reanalyses, atmospheric mesonets,
lightning-detection and air-quality networks, national avalanche services---
densely cover the \emph{meteorological} state. What none of them holds is the
quantity $\diff$ isolates: the intrinsic difficulty of a condition for a skilled
human activity, separated from the population skill mix that otherwise confounds
it. Suitability, as conventionally fit, is a blend of difficulty and clientele;
$\diff$ is the difficulty term alone. Because it is anchored to a physical curve
(Section~\ref{sec:model}), it is expressed in interpretable units and is
\emph{joinable} to those same meteorological archives: for any site and condition,
the atlas adds a difficulty coordinate the archives lack.

\subsection{A measurable construct}

The consequence is conceptual as much as practical. Two spots with identical
fitted suitability curves can have very different intrinsic difficulty, masked by
different clienteles; until the skill term is identified, they are
indistinguishable. Inverse Suitability makes them distinguishable, and in doing so
treats activity suitability---for the first time, to our knowledge---as a formal,
measurable latent construct rather than an expert convention. The atlas is the
artifact of that construct: a difficulty rating decoupled from who happens to be
present, defined precisely enough to be estimated, reproduced, and audited.

\section{Governance: reproducibility and privacy}
\label{sec:governance}

A latent model estimated from human behavioural data must be reproducible enough
to audit and private enough to respect the people in it. The two requirements
pull in opposite directions---auditability wants to retain data, privacy wants to
delete it---and the design resolves them by separating what is anonymous from what
is personal.

\subsection{Reproducible by construction}

Each fit is pinned to a content hash of the exact cohort it consumed (a SHA-256 of
the canonicalised outcome table). A cited estimate---a difficulty function or a
rider's skill posterior from cohort hash $\mathtt{abc123\ldots}$---can be
re-derived by re-running the estimator on the cohort that hash references, without
the authors' cooperation. Combined with the deterministic marginal-maximum-
likelihood estimator (Section~\ref{sec:mml}), this makes every published number
independently checkable: same cohort hash and same seed yield the same posterior.
Storing fits append-only, keyed by cohort hash, additionally turns a rider's skill
estimates over successive cohorts into a trajectory---a signal for whether skill
estimates are stable or drift after a regime change.

\subsection{Privacy: the anonymous--personal split}

The two output classes have different privacy status, and the split is the crux.
\begin{itemize}
  \item The \textbf{difficulty atlas} $\hat\diff(\metric,\site)$ is a property of
    a \emph{place}, not a person. It carries no per-rider information and is
    therefore anonymous; it survives deletion requests, in the same way that
    aggregate statistics do.
  \item The \textbf{skill posteriors} $\hat\skill_\rider$ are per-person
    attributes and are treated as sensitive: on a rider's opt-out, every skill
    estimate for that rider is hard-deleted across all cohort versions. Deletion
    is a fan-out over the append-only history, not merely a removal of the latest
    value.
\end{itemize}
This separation is what lets the atlas remain a durable, publishable scientific
artifact while the personal layer remains fully erasable. An anonymised public
export of the atlas---generalised in space and lagged in time, released only for
cells with enough distinct contributors---follows the same governance as other
anonymised environmental exports and does not depend on retaining any individual's
data.

\subsection{Skill is an input, not an inference}

Finally, a boundary that is as much a governance choice as a modelling one:
skill-conditioned scoring (Section~\ref{sec:model}) takes the skill level as an
\emph{explicit, caller-provided} input. The score for ``expert'' conditions is the
expert curve; the system does not infer a person's skill from their identity in
order to score them. This keeps the objective, physics-anchored difficulty curve
separate from any automatic profiling of individuals---skill-conditioned physics,
not personalisation---and it means the difficulty atlas can be built and used
without ever attaching a skill estimate to a named person.

\section{Discussion and applications}
\label{sec:discussion}

\paragraph{Skill-conditioned decision support.} The most direct application is
scoring that adapts to a declared skill level. Because the difficulty term is
shared and only the skill offset changes (Section~\ref{sec:model}), a single
identified model yields a family of suitability surfaces---beginner through
expert---at no additional fitting cost, each falling back invisibly to the
population curve where the latent structure is not identified. The result is a
verdict that answers ``hard for whom,'' not just ``hard on average.''

\paragraph{An intrinsic-risk primitive for underwriting.} Parametric insurance for
weather-dependent leisure needs to price the intrinsic risk of a site, not the
muddled suitability of whoever currently frequents it. The difficulty atlas is a
natural credibility primitive for such pricing: a difficulty coordinate per site
and condition, anchored to physics, reproducible from a cohort hash, and
independent of the clientele that a suitability curve would otherwise bake in. We
note the connection without building it here.

\paragraph{Limitations.} The paper's central limitation is deliberate and stated
plainly: its validation is a \emph{synthetic recovery study}, not an empirical
evaluation on real riders. We show that when data are generated from the model,
the estimator recovers the ground truth; we do \emph{not} show that real
behavioural outcomes follow the model, because no real cohort is analysed here.
Several modelling simplifications accompany this scope. The discrimination $a$ is
a single scalar rather than a function of the condition; difficulty is modelled
one physical dimension at a time rather than jointly over the full condition
vector (wind${\times}$wave${\times}$cold), which would strengthen realism at the
cost of harder identification; and estimation is point-based marginal maximum
likelihood, so the reported skills and difficulties are point estimates rather
than full posteriors. Finally, identification requires a connected
rider${\times}$condition graph (Section~\ref{sec:graph}); in real deployments many
site${\times}$condition cells will not meet that bar and will---correctly---return
the single-curve fallback rather than a latent fit.

\paragraph{Future work.} Three extensions follow directly. First, \emph{empirical
validation}: fit the model to real field cohorts as they accrue, and test the
model's structural assumptions (for instance, whether the marginal identity of
Section~\ref{sec:marginal} holds on real data, and whether the held-out skill
gain persists out of sample). Second, \emph{full Bayesian inference} with a
Gaussian-process difficulty, giving calibrated uncertainty bands on both the atlas
and the skill posteriors. Third, \emph{multi-dimensional difficulty}, jointly
identifying $\diff$ over the condition vector, which turns the per-dimension atlas
into a genuine multi-driver difficulty surface. Each is compatible with the
identification and governance framework established here.

\section{Conclusion}
\label{sec:conclusion}

Suitability scores conflate two things a decision-maker must separate: how hard a
condition is, and how skilled the person facing it is. We showed that a
continuous-item Item Response Theory model---$P(y{=}1) = \sigmoid(a(\skill_\rider -
\diff(\metric,\site)))$ with a smooth, physics-anchored difficulty function---
identifies both from binary behavioural outcomes alone. The model is strictly more
general than a single suitability curve, which it recovers exactly on marginalising
skill; it is identified when the rider${\times}$condition incidence graph is
connected, and it fails safely to the single curve when it is not. A synthetic
recovery study confirms the estimator recovers latent skill at correlation $0.96$,
locates the difficulty minimum within $\pm 3$ units of ground truth, and improves
held-out Brier Skill Score by $+0.33$ over the expert-curve baseline---all
reproducible from one command on synthetic data, with no proprietary observations.

Beyond the estimator, the recovered difficulty function defines a measurable,
anonymous, site-level construct---an intrinsic difficulty atlas---that existing
meteorological networks do not capture and that is joinable to them. Establishing
this construct, and the conditions under which it can be identified and governed,
is the paper's lasting contribution. What remains is empirical: to fit the model
to real cohorts as they accumulate and test whether human behaviour in the field
follows the structure that, in simulation, it recovers so cleanly.

\section*{Code and data availability}
\addcontentsline{toc}{section}{Code and data availability}

\sloppy
The \LaTeX{} source of this paper and a self-contained reproducibility package---
the \texttt{inverse-\allowbreak suitability latent-demo} command together with the table- and
figure-generation scripts---are released at
\url{https://github.com/goable-io/inverse-suitability} under the MIT licence
(code) and CC~BY~4.0 (text). The package reproduces every value in
Table~\ref{tab:recovery} and Figure~\ref{fig:atlas} from synthetic data with a
single command (\texttt{make repro}); no proprietary observations are required or
included. The production calibration engine, the real behavioural-outcome cohorts,
and any fitted difficulty atlases are \emph{not} part of this release---this paper
and its package concern the methodology and its synthetic validation only.

\appendix
\section{Estimation details}
\label{app:estimation}

\paragraph{Marginal maximum likelihood.} Writing $\Phi = (\diff, a)$ for the
structural parameters and $p(\skill) = \mathcal{N}(0,1)$ for the population skill
density, the marginal likelihood of rider $\rider$'s outcomes
$\{(\metric_{ri}, y_{ri})\}_i$ is
\begin{equation}
  L_\rider(\Phi) = \int \prod_i
    \sigmoid\!\big(a(\metric_{ri})(\skill - \diff(\metric_{ri}))\big)^{y_{ri}}
    \big(1 - \sigmoid(\cdot)\big)^{1 - y_{ri}}\; p(\skill)\, d\skill .
\end{equation}
The EM algorithm alternates an E-step, which forms the posterior over each rider's
skill on a fixed quadrature grid, and an M-step, which maximises the expected
complete-data log-likelihood over $\Phi$. The full log-likelihood is
$\sum_\rider \log L_\rider(\Phi)$.

\paragraph{Gauss--Hermite quadrature.} The skill integral is evaluated by
Gauss--Hermite quadrature \citep{golub1969}: nodes $\{z_q\}$ and weights
$\{w_q\}$ approximate
\begin{equation*}
  \int f(\skill)\,\mathcal{N}(\skill;0,1)\,d\skill
    \;\approx\; \sum_q w_q\, f(z_q)
\end{equation*}
after the standard change of variables. The node count
trades accuracy against cost; the demonstration configuration uses a fixed grid
adequate for a unimodal skill posterior.

\paragraph{Difficulty representation and anchoring.} The difficulty function is
represented on a metric grid and evaluated by interpolation. Its prior is centred
on the physics-derived expert curve, which anchors the difficulty origin and units
(Section~\ref{sec:ident}); rare or unobserved conditions shrink toward that
anchor. Rider skills carry the $\mathcal{N}(0,1)$ population prior, which both sets
the location/scale convention and shrinks separated (all-success or all-failure)
riders to finite values.

\paragraph{Determinism.} Given a cohort and a seed, the estimator is
deterministic---no stochastic sampling enters the point estimate---so results are
bit-for-bit reproducible.

\section{Data-generating process and reproducibility}
\label{app:repro}

\paragraph{Generator.} The reference cohort is drawn as in Section~\ref{sec:dgp}:
$80$ riders, $30$ outcomes each; skills $\skill_\rider \sim \mathcal{N}(0,1)$;
conditions uniform on $[3,35]$ knots; ground-truth difficulty
\eqref{eq:dgp-delta} with optimum $\metric^\star = 18$~kn, width $w = 8$, floor
$f = -1$; true discrimination $a_{\text{true}} = 2.0$; cohort seed fixed. Outcomes
are Bernoulli draws from the model~\eqref{eq:model}. These are the ground-truth
inputs to the \emph{generator} and are not provided to the fit.

\paragraph{One-command reproduction.} The results in Table~\ref{tab:recovery} are
produced by
\begin{quote}
  \verb|uv run inverse-suitability latent-demo --out DIR|
\end{quote}
which fits the cohort and writes \texttt{stats.json}. Under
\texttt{reproducibility/}, the paper's results table and inline macros are
rendered from that JSON by a small script (\texttt{gen\_results\_tex.py}), and the
difficulty-atlas figure is rendered from the same package's difficulty and link
functions (\texttt{gen\_figure.py}); \texttt{make repro} runs the whole
chain and rebuilds the PDF. Because the reported numbers are generated from the
code rather than transcribed, they cannot drift from it. The run is deterministic,
completes in under a minute, and uses no proprietary data and no network access.

\paragraph{Recovered versus ground-truth values.} For the avoidance of doubt: the
values in Table~\ref{tab:recovery} are the estimator's \emph{recovered} outputs
(for example, the difficulty minimum recovered at
$\hat\metric^\star = \difficultyMinKnots$~kn). The value $\metric^\star = 18$~kn is
the \emph{ground truth} supplied to the generator. The recovery claim is precisely
that the former lands within $\pm 3$~kn of the latter.

\bibliographystyle{plainnat}
\bibliography{refs}

\end{document}